\documentclass[12pt]{article}
\usepackage{epsfig,amsmath}

\title{Rotating and counterrotating relativistic thin disks\\ as
sources of stationary electrovacuum spacetimes}

\author{Gonzalo Garc\'\i a-Reyes
\thanks{e-mail: ggarcia@utp.edu.co}   \\
{\it Universidad  Tecnol\'ogica de Pereira, Departamento de 
F\'{\i}sica }  \\
{\it A. A. 97, Pereira, Colombia } \\
\and
Guillermo A. Gonz\'{a}lez
\thanks{e-mail: guillego@uis.edu.co} \\
{\it Escuela de F\'{\i}sica, Universidad Industrial de Santander}	\\
{\it A.A. 678, Bucaramanga, Colombia}  }

\date{ }

\begin{document}

\maketitle

\begin{abstract}
A detailed study is presented of the counterrotating model (CRM) for
electrovacuum stationary axially symmetric relativistic thin disks of infinite
extension without radial stress, in the case when the eigenvalues of the
energy-momentum tensor of the disk are real quantities, so that there is not
heat flow. We find a general constraint over the counterrotating tangential
velocities needed to cast the surface  energy-momentum tensor of the disk as
the superposition of two counterrotating charged dust fluids. We then show
that, in some cases, this constraint can be satisfied if we take the two
counterrotating  tangential velocities as equal and opposite or by taking the
two counterrotating streams as circulating along electro-geodesics. However, we
show that, in general, it is not possible to take the two counterrotating
fluids as circulating along electro-geodesics nor take the two counterrotating
tangential velocities as equal and opposite. A simple family of models of
counterrotating charged disks based on the Kerr-Newman solution are considered
where we obtain some disks with a CRM well behaved. We also show that the disks
constructed  from the Kerr-Newman solution can be interpreted, for  all the
values of parameters, as a matter distribution with currents and purely
azimuthal pressure without heat flow. The models are constructed using the
well-known ``displace, cut and reflect'' method extended to solutions of vacuum
Einstein-Maxwell equations. We obtain, in all the cases, counterrotating
Kerr-Newman disks that are in agreement with all the energy conditions.

\end{abstract}

\section{Introduction}

Several methods are known to exactly solve the Einstein  and  Einstein-Maxwell
field equations, or to generate new  exact  solutions from simple known
solutions \cite{KSHM}.  However, the above mentioned methods in general lead
to  solutions without a clear physical interpretation or to  solutions that
depend on many parameters without a clear  physical meaning. Accordingly, it is
of importance to have some appropriate procedures to obtain physical
interpretations of these exact solutions. So, in the past  years such
procedures have been developed for static and  stationary  axially symmetric
solutions in terms of thin  and, more  recently, thick disk models.

Stationary or static axially symmetric exact solutions of   Einstein  equations
describing relativistic thin disks are of  great astrophysical importance since
they can be used  as  models of certain stars, galaxies and accretion disks.
These were first studied by Bonnor and Sackfield \cite{BS}, obtaining
pressureless static disks, and then by Morgan and Morgan, obtaining static
disks with and without radial  pressure \cite{MM1,MM2}. In connection with
gravitational collapse, disks were first studied by Chamorro, Gregory, and 
Stewart. Also thin disks with radial tension were considered \cite{GL1}.  
Several classes of exact solutions of the  Einstein field  equations
corresponding to static thin disks with or  without radial pressure have been
obtained by different authors [\ref{bib:LP} - \ref{bib:GE}].  

Rotating  thin disks that can be considered as a source of a Kerr metric were
presented by  Bi\u{c}\'ak and  Ledvinka \cite{BL}, while rotating disks with
heat flow were studied by Gonz\'alez and Letelier \cite{GL2}. The nonlinear 
superposition of a disk and a black hole was first obtained by Lemos and
Letelier \cite{LL1}. Perfect fluid disks with halos were studied by Vogt and
Letelier \cite{VL1}. The stability of some general relativistic thin disks
models using a first order perturbation of the energy-momentum tensor was
investigated by Ujevic and Letelier \cite{UL1}.  

Gonz\'alez and Letelier \cite{GL3}  constructed models of   static relativistic
thick disks in various coordinate systems. Although the disks have constant
thickness, the matter density decreases rapidly with radius and the $z$ 
coordinate, and in principle they also can be used to represent both the disk
part and the central bulges of galaxies. Also Vogt and Letelier \cite{VL3} 
considered more  realistic three-dimensional models for the gravitational field
of Galaxies in the General Relativistic context.  Essentially they formulate
the General Relativistic  versions in isotropic coordinates of the
potential-density pairs deduced by Miyamoto and Nagai \cite{MN1,MN2} and Satoh
\cite{Sat}.

Disk sources for stationary axially symmetric spacetimes  with  magnetic fields
are also of astrophysical importance  mainly in the study of neutron stars,
white dwarfs and  galaxy formation. Although disks with electric fields do not
have clear  astrophysical importance, their study may be  of interest in the
context of exact solutions. Thin disks have been discussed as sources for
Kerr-Newman fields \cite{LBZ}, magnetostatic  axisymmetric fields \cite{LET1},
conformastationary metrics \cite{KBL},  while models of electrovacuum static
counterrotating dust disks  were presented in \cite{GG1}. Charged perfect fluid
disks were also studied by Vogt and Letelier \cite{VL2}, and  charged perfect
fluid disks as sources of  static and  Taub-NUT-type spacetimes by Garc\'\i
a-Reyes and Gonz\'alez \cite{GG2,GG3}. 

In all the above cases, the disks are obtained by an ``inverse problem''
approach, called by Synge the ``{\it g-method}'' \cite{SYN}. The method works
as follows: a solution of the vacuum Einstein equations is taken, such that
there is a discontinuity in the derivatives of the metric tensor on the plane
of the disk,  and the energy-momentum tensor is obtained from the Einstein
equations. The physical properties of the matter  distribution are then studied
by an analysis of the surface energy-momentum tensor so obtained. Another
approach to  generate disks is by solving the Einstein equations given a source
(energy-momentum tensor). Essentially, they are obtained by solving a
Riemann-Hilbert problem and are highly nontrivial [\ref{bib:NM} -
\ref{bib:KLE4}].  
A review of this kind of disks solutions to the Einstein-Maxwell
equations was presented by Klein in \cite{KLE5}.

Now, when the inverse problem approach is used for static electrovacuum
spacetimes, the energy-momentum tensor is diagonal and its analysis is direct
and, except for the dust disks, all the obtained disks have anisotropic sources
with azimuthal stress  different from the radial stress. On the other hand,
when the considered spacetime is stationary, the obtained energy-momentum
tensor is non-diagonal and the  analysis of its physical content is more
involved and, in general, the obtained source is not only anisotropic but with
nonzero heat flow. Due to this fact, there are  very few  works about of
stationary electrovacuum disks and they are limited to disks  obtained with
solutions that lead to disks without heat flow \cite{LBZ,GG3}.

The necessary condition to obtain a thin disk without heat flow is that   the
eigenvalues of the energy-momentum tensor must be real quantities, which can be
only  for very few known electrovacuum solutions. In \cite{GG3} we consider a
Taub-NUT type solution such that the energy-momentum tensor can be written as
an upper right triangular matrix, so that the diagonalization is trivial and
the eigenvalues are real quantities. However, the obtained disks are not really
rotating disks since the spatial components of their velocity vectors are zero
with respect to the coordinates and so the disks are ``locally statics''. The
first true rotating electrovacuum thin disks were obtained by Ledvinka,
Bi\u{c}\'{a}k, and \u{Z}ofka \cite{LBZ} by applying the ``displace, cut and
reflect'' method to the Kerr-Newman solution. The so obtained disks have no
radial pressure and no heat flow. However, the authors do not show if  the
eigenvalues of the momentum-energy tensor are real quantities for all the
values of the parameters, is that, if these disks can always be interpreted as
a matter distribution with currents and purely azimuthal pressure,  or if there
are some case where can exist nonzero heat flow (complex eigenvalues).

The above disks can also be interpreted as made of two counterrotating streams
of moving charged particles, as was also indicated in \cite{LBZ}. Now, in order
to do this interpretation, the counterrotating tangential velocities of the two
streams must to satisfy a constraint, which in general is not satisfied for
disks obtained from generic stationary electrovacuum solutions. In \cite{LBZ}
the authors take the two counterrotating streams as circulating along
electro-geodesics, but they do not show if such decomposition can be done. In
addition, as we will show in this paper, in general the electro-geodesics
motion do not agree with the above mentioned constraint and so it is necessary
to consider another possibility for the complete determination of the
counterrotating tangential velocities. Another possibility, also commonly 
assumed, is to take the two counterrotating velocities as equal and opposites
but, as we will show, in general the counterrotating velocities are not
completely determined  by the constraint, so that the corresponding
interpretation as two counterrotating  streams is not possible.

The above interpretation is obtained by means of the Counterrotating Model
(CRM) in which the energy-momentum tensor of the source is expressed as the
superposition of two counterrotating perfect fluids. Now, even though this
interpretation can be seen as merely theoretical, there are observational
evidence of disks made of streams of rotating and counterrotating matter (see,
for instance, \cite{RGK,RFF,BER}). These disks are made of stars and gas so
that they are disks with pressure.  Nevertheless,  as is suggested in
\cite{RFF}, the preexisting galaxies have a component  originally constituted
mainly by a gas free stellar disk, i.e., collisionless matter or dust. A
detailed study of the CRM for generic relativistic static thin disks was
presented in \cite{GE} for the vacuum case, whereas the extension for static
electrovacuum disks was presented in \cite{GG1,GG2}. On the other hand, the CRM
for stationary thin disks has not been completely developed, neither for the
vacuum case, and only a preliminary version of it was presented in \cite{GL2}
for the case of stationary thin disks without heat flow and with positive
radial stress (pressure).

The purpose of the present paper is twofold. In first instance, we present a
detailed analysis of the energy-momentum tensor and the surface current density
for electrovacuum stationary axially symmetric relativistic thin disks of
infinite extension without radial stress, in the case when the energy-momentum
tensor of the disks can be diagonalized, so that there is not heat flow. And,
in the second place, we present the complete study of the Counterrotating Model
for these stationary thin disks. The paper is  structured as follows. In Sec.
II  we present a summary of the procedure to obtain  models of  rotating thin
disks  with a purely azimuthal pressure  and  currents, using the well-known
``displace, cut  and reflect'' method extended to solutions of 
Einstein-Maxwell equations,  in the case  when the  eigenvalues of the
energy-momentum tensor of the disk  are real quantities.  In particular, we
obtain expressions for the surface  energy-momentum tensor  and the surface
current density of the disks.   

In Sec. III the disks are interpreted in terms of  the counterrotating model
(CRM). We find the general constraint over the counterrotating  tangential
velocities needed to cast the surface  energy-momentum tensor of the disk as
the superposition of two counterrotating  charged dust fluids.  We then show 
that this constraint can be satisfied if we take the two counterrotating 
tangential velocities as equal and opposite as well as by taking the two
counterrotating streams as circulating along electro-geodesics.  However, we
show that, in general, it is not possible to take the two counterrotating
fluids as circulating along electro-geodesics nor take  the two counterrotating
tangential velocities as equal and opposite.  We also find explicit expressions
for the energy densities, current densities and velocities of the two
counterrotating fluids. 

In the following section, Sec. IV, we consider a family of models of 
counterrotating charged dust disks based on the Kerr-Newman metric, perhaps the
only simple electrovacuum solution that lead to stationary thin disks without
heat flow. We show that for Kerr-Newman fields the eigenvalues of the
energy-momentum  tensor of the disks are always real quantities, for all the
values of the parameters, and so they do not present heat flow in any case.  We
also analyze the CRM for these disks and study the  tangential velocities,
energy and electric charge densities  of both  streams when the two fluids move
along electrogeodesics and when they move with equal and opposite  velocities.
Also the stability against radial  perturbation is analyzed in both of the
cases. Finally, in Sec. V, we summarize our main results.   

\section{Electrovacuum rotating relativistic thin disks}

A sufficiently general metric for our purposes can be  written as the
Weyl-Lewis-Papapetrou line element \cite{KSHM},   
\begin{equation}
ds^2 = - \ e^{2 \Psi} (dt + {\cal W} d\varphi)^2 \ +  \ e^{- 2 \Psi}[ r^2 
d\varphi^2 + e^{2 \Lambda} (dr^2 + dz^2)], \label{eq:met}
\end{equation}
where we use for the coordinates the notation $(x^0,x^1,x^2,x^3) =
(t,\varphi,r,z)$, and $\Psi$, ${\cal W}$, and $\Lambda$ are functions of $r$
and $z$ only.  The vacuum Einstein-Maxwell  equations,  in  geometric units in
which $8 \pi G = c = \mu _0 =  \varepsilon _0 = 1$,  are given by 
\begin{subequations}\begin{eqnarray}
&   &   G_{ab} \  =  \ T_{ab},  \label{eq:einmax1}    \\
&   &    F^{ab}_{ \ \ \ ; b} \ =  \ 0 \label{eq:einmax2},     
\end{eqnarray}\end{subequations}
with
\begin{subequations}\begin{eqnarray}
T_{ab} \  &=&  \ F_{ac}F_b^{ \ c} - \frac{1}{4} g_{ab}  
F_{cd} F^{cd}, \label{eq:et}  \\
F_{ab} \ &=&  \  A_{b,a} -  A_{a,b},
\end{eqnarray}\end{subequations} 
where $A_a = (A_t, A_\varphi, 0, 0)$ and the electromagnetic  potentials  $A_t$
and $A_\varphi$ are also functions of  $r$ and $z$  only.

For the metric (\ref{eq:met}), the Einstein-Maxwell equations are equivalent to
the system \cite{E2}
\begin{subequations}\begin{eqnarray}
&& \qquad \qquad \nabla \cdot [ r^{-2}f( \nabla A_\varphi - {\cal W} \ \nabla
A_t ]=0, \label{em11} \\
&&  \qquad \quad \nabla \cdot [ f^{-1} \nabla A_t + r^{-2} f {\cal W} ( \nabla
A_\varphi - {\cal W} \ \nabla A_t ]=0, \label{em12} \\
&& \qquad \nabla \cdot [r^{-2} f^{2} \nabla {\cal W} - 2 r^{-2}
f A_t (\nabla A_\varphi - {\cal W} \ \nabla A_t)] = 0 , \label{em13} \\
&&f \nabla^{2} f  = \nabla f \cdot \nabla f - r^{-2} f^{4} \nabla{\cal W}
\cdot  \nabla{\cal W} + f \nabla A_t \cdot \nabla A_t  \nonumber \\
&& \qquad \qquad + \ r^{-2} f^{3} (\nabla A_\varphi - {\cal W} \nabla A_t)
\cdot (\nabla A_\varphi - {\cal W} \nabla A_t) , \label{em14} \\
\Lambda_{,r} &=& r (\Psi_{,r}^2 - \Psi_{,z}^2 )
- \frac 1 {4r} ({\cal W}_{,r}^2 -{\cal W}_{,z}^2)e^{4 \Psi }
- \frac 1 {2r} (r^2 e^{-2 \Psi }-{\cal W}^2e^{2 \Psi }) 
(A_{t,r}^2-A_{t,z}^2) \nonumber  \\
&&+ \ \frac 1 {2r}  (A_{\varphi,r}^2-A_{\varphi,z}^2) e^{2 \Psi }
-\frac{1}{r}{\cal W}(A_{\varphi,r} A_{t,r}-
A_{\varphi,z} A_{t,z})e^{2 \Psi}, \\
\Lambda_{,z} &=&  2r \Psi_{,r} \Psi_{,z} - \frac{1}{2r} 
{\cal W}_{,r} {\cal W}_{,z} e^{4 \Psi} -\frac{1}{r}(r^2
e^{-2 \Psi} -{\cal W}^2 e^{2 \Psi})A_{t,r} A_{t,z} \nonumber \\
&& + \ \frac{1}{r}A_{\varphi,r} A_{\varphi,z} e^{2 \Psi} 
 -\frac{1}{r}{\cal W}(A_{\varphi,r} A_{t,z}+
A_{\varphi,z} A_{t,r})e^{2 \Psi},  \label{eq:lamz}
\end{eqnarray}\end{subequations}
where $\nabla$ is the  standard  differential 
operator in cylindrical coordinates and $f = e^{2 \Psi}$. 

In order to obtain a solution of (\ref{eq:einmax1}) - (\ref {eq:einmax2}) 
representing a thin disk at $z=0$, we assume  that the components of the metric
tensor are continuous across the disk, but with first  derivatives
discontinuous on the  plane $z=0$, with  discontinuity functions 
\begin{eqnarray}
b_{ab} \ &=& g_{ab,z}|_{_{z = 0^+}} \ - \ g_{ab,z}|_{_{z = 0^-}} \ = \ 2 \
 g_{ab,z}|_{_{z = 0^+}}.                  
\end{eqnarray}
Thus, by using the distributional approach \cite{PH,LICH,TAUB} or the junction
conditions on the extrinsic curvature of thin shells \cite{IS1,IS2,POI}, the
Einstein-Maxwell equations yield an  energy-momentum tensor $T_{ab}=
T^{\mathrm{elm}}_{ab} +  T^{\mathrm {mat}}_{ ab}$,  where 
$T^{\mathrm{mat}}_{ab} = Q_{ab} \ \delta (z)$, and a  current density $J_a =
j_a  \delta (z)= - 2 e^{2 (\Psi - \Lambda)} A_{a,z} \delta (z)$, where $\delta
(z)$ is the  usual Dirac function with support on  the disk.  $T^{\mathrm {
elm}}_{ab}$ is the electromagnetic tensor defined in Eq. (\ref{eq:et}), $j_a$
is the current density on the plane  $z=0$, and 
$$
Q^a_b = \frac{1}{2}\{b^{az}\delta^z_b - b^{zz}\delta^a_b +  g^{az}b^z_b -
g^{zz}b^a_b + b^c_c (g^{zz}\delta^a_b - g^{az}\delta^z_b)\}
$$
is the distributional energy-momentum tensor. The ``true''  surface
energy-momentum tensor (SEMT) of the  disk,  $S_{ab}$, and the ``true'' surface
current density,  $\mbox{\sl j}_a$, can be obtained through the
relations  	
\begin{subequations}\begin{eqnarray}
S_{ab} &=& \int T^{\mathrm {mat}}_{ab} \ ds_n \ = \ e^{  \Lambda - \Psi} \ 
Q_{ab} \ ,   \\
\mbox{\sl j}_a  &=& \int J_{a}  \ ds_n \ = \ e^{ \Lambda -  \Psi} \ j_a \ , 
\end{eqnarray}\end{subequations}
where $ds_n = \sqrt{g_{zz}} \ dz$ is the ``physical  measure'' of length in the
direction normal to the disk.

For the metric (\ref{eq:met}), the nonzero components of  $S_a^b$ are
\begin{subequations}\begin{eqnarray}
&S^0_0 &= \ \frac{e^{\Psi - \Lambda}}{ r^2} \left [ 2 r^2 (\Lambda,_z - \ 2
\Psi,_z) - \ e^{4\Psi} {\cal W} {\cal W},_z \right ] , \label{eq:emt1}  \\
&S^0_1 &= \ - \frac{e^{\Psi - \Lambda}}{ r^2} \left [  4 r^2 {\cal W}  \Psi,_z
+ \ ( r^2 + \ {\cal W}^2 e^{4\Psi} ) {\cal W},_z \right ] , \label{eq:emt2}  \\
&S^1_0 &= \ \frac{e^{\Psi - \Lambda}}{ r^2}  \left[ e^{4\Psi}  {\cal W},_z
\right ], \label{eq:emt3} \\
&S^1_1 &= \ \frac{e^{\Psi - \Lambda}}{ r^2} \left [ 2 r^2 \Lambda,_z + \
e^{4\Psi} {\cal W} {\cal W},_z \right ] ,  \label{eq:emt4} 
\end{eqnarray}\end{subequations} 
and the nonzero components of the surface current density  $\mbox{\sl j}_a$
are   
\begin{subequations}\begin{eqnarray}
& \mbox{\sl j}_t &= \ -2 e^{\Psi - \Lambda} A _{t,z} , 
\label{eq:corelec}   \\
&\mbox{\sl j}_{\varphi} &= \ -2 e^{\Psi - \Lambda} 
A _{\varphi ,z},  \label{eq:cormag} 
\end{eqnarray}\end{subequations}
where all the quantities are evaluated at $z = 0^+$.

These disks are essentially of  infinite extension. Finite disks can be
obtained  introducing oblate  spheroidal coordinates,  which are  naturally
adapted to a disk source, and  imposing  appropriate boundary conditions. These
solutions, in the vacuum and static case, correspond to the Morgan and Morgan
solutions \cite{MM1}. A more general class of  solutions representating finite
thin disks can be constructed  using a method based on the use of conformal
transformations and solving a boundary-value problem  \cite{MM2,
CHGS,GL1,GE,GG2,GG3}. 
 
Now, in order to analyze the matter content of the disks is   necessary to
compute the eigenvalues and eigenvectors of  the energy-momentum tensor. The
eigenvalue problem for the SEMT (\ref{eq:emt1}) - (\ref{eq:emt4})
\begin{equation}
S^a_b \ \xi^b \ = \lambda \ \xi^a,
\end{equation}
has the solutions
\begin{equation}
\lambda_\pm \ = \ \frac{1}{2} \left( \ T \pm \sqrt{D} \  \right) , 
\end{equation}
where
\begin{equation}
T = S^0_0 \ + \ S^1_1 \ , \quad D = ( S^1_1 - S^0_0 )^2 +  4 \ S^0_1 \ S^1_0 ,
\end{equation}
and $\lambda_r = \lambda_z = 0$. For the metric  (\ref{eq:met})    
\begin{subequations}\begin{eqnarray}
D  & = & 4 \frac{e^{2(\Psi - \Lambda)}}{r^2} (4r^2  \Psi_{,z}^2-{\cal W}_{,z}^2
e^{4\Psi} ) = A^2 - B^2, \\
T &=& 4 e^{\Psi - \Lambda}(\Lambda_{,z} - \Psi_{,z}),
\end{eqnarray}\end{subequations}  
where 
\begin{equation}
A=4 \Psi_{,z}e^{\Psi - \Lambda} , \ \ \ \ B = \frac 2 r   {\cal W}_{,z}
e^{3\Psi -\Lambda}.
\end{equation}
The corresponding eigenvectors are
\begin{equation}\begin{array}{ccl}
\xi^a_\pm & = &  (\ \xi^0_\pm  , \ \xi^1_\pm  , \ 0, \ 0 ), \\
	&	&	\\
X^a & = & e^{U - \Lambda} ( 0, 0, 1, 0 ),	\\
	&	&	\\
Y^a & = & e^{U - \Lambda} ( 0, 0, 0, 1 ),
\end{array}\label{eq:tetrad}\end{equation}
with
\begin{equation}
g (\xi_\pm, \xi_\pm ) = 2  N_\pm e^{2\Psi} \left (\frac  {\xi^0_\pm }{S^0_0 -
S^1_1 \pm \sqrt D } \right )^2    , 
\end{equation}
where
\begin{equation}
 N_\pm  = \sqrt { D} (-\sqrt { D} \pm A). \label{eq:norm}
\end{equation}

We only consider the case when $D \geq 0$, so that the two eigenvalues
$\lambda_\pm$ are real  and different and the two eigenvectors are orthogonal, 
in such a way that one of them is timelike and the other  is spacelike. Since
$|A|\geq \sqrt { D}$, from (\ref{eq:norm}) follows  that when $A>0$ the
negative sign corresponds to the  timelike eigenvector while the  positive sign
to the  spacelike eigenvector. When  $A<0$ we have the opposite case. So  the
function $\Psi _{,z}$ determines the sign of the norm.

Let $V^a$ be the timelike eigenvector, $V_aV^a = -1$, and 
$W^a$ the spacelike  eigenvector, $W_aW^a = 1$.
In terms of the  orthonormal tetrad or comoving observer  ${{\rm e}_{\hat a}}^b
= \{ V^b , W^b , X^b , Y^b \}$,  the SEMT and the surface electric current
density may be decomposed as  
\begin{subequations}\begin{eqnarray}
S_{ab} \ &=& \ \epsilon V_a V_b + p_\varphi W_a W_b , \label{eq:emtcov} \\
\mbox{\sl j}_a \ &=& \ \mbox{\sl j}^{\hat 0} V_a + 
\mbox{\sl j}^{\hat 1} W_a,  \label{eq:ja} 
\end{eqnarray}\end{subequations}
where
\begin{equation}
\epsilon \ = \ - \lambda_{\pm}, \quad \quad  p_\varphi
 \ = \ \lambda_{\mp},
\label{eq:enpr}
\end{equation}
are, respectively, the surface energy density, the  azimuthal pressure, and
\begin{equation}
\mbox{\sl j}^{\hat 0} = - V^a \mbox{\sl j}_a,  \quad 
\mbox{\sl j}^{\hat 1}  =  W^a \mbox{\sl j}_a,  \label{eq:djs}
\end{equation}
are the surface electric charge density and the azimuthal current  density  of the disk 
measured by this observer.  In (\ref{eq:enpr}) the sign is chosen according to which  is the
timelike  eigenvector and which is the spacelike eigenvector.
However, in order to satisfy  the strong energy condition
$\varrho= \epsilon + p_\varphi  \geq 0$, where $\varrho$ is
the effective Newtonian density, we must choose 
$\xi _-$ as the timelike eigenvector and  $\xi _+$ as the
spacelike eigenvector.
These condition characterizes a disk made of matter with 
the usual gravitational attractive property.
Consequently  $\Psi _{,z}$ must be taken positive.
So we have
\begin{equation}
\epsilon \ = \ - \lambda_-, \quad \quad  p_\varphi
 \ = \ \lambda_+,
\end{equation}
and
\begin{subequations}\begin{eqnarray}
V^0 &=& \frac{\nu e^{-\Psi}} { \sqrt{- 2 N_-}} (S^0_0 - S^1_1 - \sqrt D),  \\
V^1 &=&  \frac{2 \nu  e^{-\Psi}} {   \sqrt{- 2 N_-}} S^1_0 ,
\end{eqnarray}\end{subequations}
where $ \nu = \pm $ so that the sign is chosen  according  to  the causal
character of the timelike eigenvector  (observer's four-velocity),  
\begin{subequations}\begin{eqnarray}
W^0 &=& \frac{2} {\sqrt{2 M}}S^0_1,  \\
W^1 &=& \frac{1} { \sqrt{2 M}} (S^1_1-S^0_0 + \sqrt{ D} ),
\end{eqnarray}\end{subequations}
where
\begin{equation}
M = \sqrt{ D} \left \{ g_{11} \sqrt{ D}  +2r{\cal W} B
+ (r^2 e^{-2\Psi} + {\cal W}^2  e^{2\Psi} )A  \right \}.
\end{equation}


\section{Counterrotating charged dust disks}

We now consider, based on Refs. \cite{GE} and \cite{GG2}, the possibility that
the SEMT $S^{ab}$ and the current density  $\mbox{\sl j}^a$ can be written as
the superposition of two counterrotating charged  fluids that circulate in
opposite directions; that is, we assume 
\begin{subequations}\begin{eqnarray}
S^{ab} &=& S_+^{ab} \ + \ S_-^{ab} \ , \label{eq:emtsum}   \\
\mbox{\sl j}^a    &=& \mbox{\sl j}_+^a + \mbox{\sl j}_-^a,  \label {eq:corsum} 
\end{eqnarray}\end{subequations}
where the  quantities on the right-hand side are, respectively, the SEMT and
the current density of the prograde and retrograde counterrotating fluids. 

Let  $U_\pm^a = ( U_\pm^0 , U_\pm^1, 0 , 0 )= U_\pm^0(1, \omega _\pm, 0 , 0 )$
be the velocity vectors of the two counterrotating fluids, where  $\omega_\pm =
U_\pm^1/U_\pm^0$ are the angular velocities of each stream.  In order to do the
decomposition  (\ref{eq:emtsum}) and (\ref{eq:corsum}) we project the velocity
vectors onto the tetrad ${{\rm e}_{\hat a}}^b$, using the relations 
\cite{CHAN}
\begin{equation}
U_\pm^{\hat a} \ = \ {{\rm e}^{\hat a}}_b U_\pm^b, \quad  \quad U_\pm^ a = \ 
{{\rm e}_{\hat b}}^a U_\pm^{\hat b} .
\end{equation}
In terms of  the tetrad (\ref{eq:tetrad}) we can write
\begin{equation}
U_\pm^a \ = \ \frac{ V^a + v_\pm W^a }{\sqrt{1 - v_\pm^2}} , \label{eq:vels}
\end{equation}
so that 
\begin{subequations}\begin{eqnarray}
&V^a &= \ \frac{\sqrt{1 - v_-^2} v_+ U_-^a - \sqrt{1 - v_+^2} v_- U_+^a}{ v_+ -
v_-}  , \label{eq:va} \\
&W^a &= \ \frac{\sqrt{1 - v_+^2} U_+^a - \sqrt{1 - v_-^2} U_-^a}{v_+ - v_-}  ,
\label{eq:wa}
\end{eqnarray}\end{subequations}
where $v_\pm = U_\pm^{\hat 1} / U_\pm^{\hat 0}$ are the tangential velocities
of the streams with respect to the tetrad.

Another quantity related with the counterrotating motion is the specific
angular momentum of a particle rotating at a radius $r$, defined as $h_\pm =
g_{\varphi a} U_\pm^a$.
This quantity can be used to analyze the stability of circular
orbits of test particles 
against radial
perturbations. The condition of stability,
\begin{equation}
\frac{d(h^2)}{dr} \ > \ 0  ,
\end{equation}
is an extension of Rayleigh criteria of stability of a fluid in rest in a
gravitational field \cite{FLU}. For an analysis of the stability
of a rotating fluid  taking into account the collective behavior of the particles see for example Refs. \cite{FHS,UL1}.

Substituting  (\ref{eq:va}) and (\ref{eq:wa}) in  (\ref{eq:emtcov}) we obtain
\begin{eqnarray}
S^{ab} & = & \frac{ F( v_- , v_- ) (1 - v_+^2) \ U_+^a U_+^b }{(v_+ - v_-)^2}
\nonumber	\\
& + & \frac{ F( v_+ , v_+ ) (1 - v_-^2) \ U_-^a U_-^b }{(v_+ -
v_-)^2}			\nonumber	\\
& - & \frac{ F( v_+ , v_- ) (1 - v_+^2)^{\frac{1}{2}} (1 - v_-^2)^{\frac{1}{2}}
( U_+^a U_-^b + U_-^a U_+^b ) }{(v_+ - v_-)^2}		\nonumber	
\end{eqnarray}
where
\begin{equation}
F( v_1 , v_2 ) \ = \  \epsilon  v_1 v_2 + p_\varphi.  \label{eq:fuu}
\end{equation}
Clearly, in order to cast the SEMT in the form (\ref{eq:emtsum}), the mixed
term must be absent and therefore the counterrotating tangential velocities
must satisfy the  following constraint
\begin{equation}
F( v_+ , v_- ) \ = \ \epsilon  v_+ v_- + p_\varphi=0  , \label{eq:liga}
\end{equation}
where we assume that $|v_\pm| \neq 1$. 

Then, assuming a given choice for the tangential velocities in agreement with
the above relation, we can write the SEMT as (\ref{eq:emtsum}) with
\begin{equation}
S^{ab} _\pm = \epsilon_\pm \ U_\pm ^a U_\pm ^b, \label{eq:sabcon}
\end{equation}
so that we have two counterrotating dust fluids with 
surface energy densities, measured in the coordinates 
frames, given by 
\begin{equation}
\epsilon_\pm =  \left[ \frac{ 1 - v_\pm^2 }{v_\mp - v_\pm}\right]   \epsilon
v_\mp , \label{eq:epcon}
\end{equation}
Thus the SEMT $S^{ab}$ can be written as the superposition of two
counterrotating dust fluids if, and only if, the constraint (\ref{eq:liga}) 
admits a solution such that $v_+ \neq v_-$. 

Similarly, substituting  (\ref{eq:va}) and (\ref{eq:wa}) in  (\ref{eq:ja}) we
can write the current density as (\ref{eq:corsum}) with
\begin{equation}
\mbox{\sl j}^a_\pm  = \sigma _\pm U_\pm ^a  \label{eq:jacon}
\end{equation}
where $\sigma _\pm$ are the  surface electric charge densities, measured in the coordinates frames, 
\begin{equation}
\sigma _{ \pm} =  \left[ \frac { \sqrt{1-v^2_\pm }} 
{  v_\pm -v_\mp} \right] (\mbox{\sl j}^{\hat 1}
- \mbox{\sl j}^{\hat 0} v_\mp  ). \label{eq:sig} 
\end{equation}
Thus, we have a disk  makes of two counterrotating charged dust fluids with
surface energy densities given by  (\ref{eq:epcon}), and surface electric charge densities
given by  (\ref{eq:sig}).


As we can see from Eqs. (\ref{eq:vels}), (\ref{eq:epcon})  and (\ref{eq:sig}),
all the main physical quantities  associated with the CRM  depend on  the
counterrotating tangential velocities $v_\pm$.   However, the constraint 
(\ref{eq:liga}) does not determine  $v_\pm$ uniquely so that we need to impose 
some additional requirement in order to obtain a complete determination of  the
tangential velocities leading  to a well defined CRM.


A possibility, commonly assumed \cite{LBZ,KLE3}, is to take the two  counterrotating streams as
circulating  along electrogeodesics. Now, if the electrogeodesic equation
admits solutions corresponding to circular orbits, we can write this equation
as
\begin{equation}
\frac 12 \epsilon _\pm g_{ab,r}U^a_\pm U^b_\pm = - \sigma _\pm F_{ra} U^a_\pm.
\label{eq:elecgeo}
\end{equation}
In terms of $\omega _\pm$ we obtain
\begin{equation}
\frac 12 \epsilon _\pm (U^0_\pm)^2 ( g_{11,r} \omega _\pm ^2  + 2g_{01,r}
\omega _\pm + g_{00,r} ) = - \sigma _{\pm} U_\pm ^0 ( A _{t,r} + A_{\varphi,r}
\omega _{\pm}).  \label{eq:elecgeo1}
\end{equation}
From (\ref{eq:emtsum}), (\ref{eq:corsum}), (\ref{eq:sabcon}), and
(\ref{eq:jacon}) we have
\begin{subequations}\begin{eqnarray}
\sigma_\pm U^0_\pm & = & \frac{\mbox{\sl j}^1 -  \omega_\mp \mbox{\sl
j}^0}{\omega_\pm - \omega_\mp}, \label{eq:sigmau0} \\
\epsilon_\pm (U^0_\pm)^2 & = & \frac{S^{01} - \omega_\mp  S^{00}}{\omega_\pm -
\omega_\mp}, \label{eq:epsilonu02} \\
\omega_\mp & = &  \frac {S^{11} - \omega_\pm S^{01} } {S^{01} - \omega _ \pm
S^{00}}, \label{eq:omegapm}
\end{eqnarray}\end{subequations}
and substituting  (\ref{eq:sigmau0}) and (\ref{eq:epsilonu02}) in 
(\ref{eq:elecgeo1}) we find  
\begin{equation}
\frac 12 (S^{01} - \omega_\mp S^{00}) ( g_{11,r} \omega _\pm ^2 + 2g_{01,r}
\omega _\pm + g_{00,r} ) =  - (\mbox{\sl j}^1 -  \omega_\mp \mbox{\sl j}^0) ( A
_{t,r} + A_{\varphi,r}  \omega _{\pm}),  \label{eq:elecgeo2}
\end{equation}
and using (\ref{eq:omegapm}) we obtain
\begin{equation}
\frac 12[(S^{01})^2 - S^{00}S^{11}] ( g_{11,r} \omega _\pm ^2 + 2g_{01,r}
\omega _\pm + g_{00,r} ) =     - [S^{01}\mbox{\sl j}^1 - S^{11}\mbox{\sl j}^0  
+  \omega_\pm (S^{01} \mbox{\sl j}^0 - S^{00}\mbox{\sl j}^1)] ( A _{t,r} +
A_{\varphi,r} \omega _{\pm}).  \label{eq:elecgeo3}
\end{equation}
Therefore we conclude that 
\begin{equation}
\omega _\pm = \frac {-T_2 \pm \sqrt{T_2^2 - T_1 T_3}} {T_1}  \label{eq:omega}
\end{equation}
with
\begin{subequations}\begin{eqnarray}
T_1 &=& g_{11,r} + 2 A_{\varphi,r} \frac{\mbox{\sl j}^0  S^{01} - \mbox{\sl
j}^1 S^{00} } { S^{01}S^{01} - S^{00} S^{11} } , \label{eq:T1} \\
T_2 &=& g_{01,r} + A_{t,r} \frac{ \mbox{\sl j}^0 S^{01}  - \mbox{\sl j}^1
S^{00} } {S^{01} S^{01} - S^{00} S^{11}} + A_{\varphi,r}  \frac{ \mbox{\sl j}^1
S^{01}  - \mbox{\sl j}^0 S^{11} } { S^{01} S^{01} - S^{00} S^{11} }, 
\label{eq:T2} \\
T_3 &=& g_{00,r}  + 2 A_{t,r} \frac{\mbox{\sl j}^1 S^{01}  - \mbox{\sl j}^0
S^{11}} {S^{01} S^{01} - S^{00} S^{11} }. \label{eq:T3}
\end{eqnarray}\end{subequations}

On the other hand, in terms of $\omega _\pm$ we get
\begin{equation}
v_{\pm} = - \left [ \frac{W_0 + W_{1} \omega _\pm} {V_0 + V_{1} \omega _\pm}
\right ], \label{eq:vrotcon}
\end{equation}
and so, by using (\ref{eq:omega}), we have that
\begin{equation}
v_+v_- = \frac{T_1 W_0^2-2T_2W_0 W_1+ T_3 W_1^2  }{T_1 V_0^2-2T_2V_0 V_1+ T_3
V_1^2 },
\end{equation}
so that, using  (\ref{eq:emtcov}), we get
\begin{eqnarray}
F(v_+,v_-) & = &\frac{32 e^{4(\Psi -\Lambda)} \Lambda _{,z}^2  ( r^2 \Lambda 
_{,z} \sqrt{D} +4r^2 \Lambda _{,z} \Psi_{,z} e^{\Psi - \Lambda} - {\cal
W}_{,z}^2 e^{5\Psi - \Lambda}  ) } {r^3(A+ \sqrt{D}) p_\varphi (S^{01}S^{01} -
S^{00} S^{11})(T_1 V_0^2-2T_2 V_0 V_1 +T_3V_1^2)} \nonumber   \\
& & \times \left [ \Lambda _{,z} - 2r \Psi_{,r} \Psi_{,z} + \frac{1}{2r} {\cal
W}_{,r} {\cal W}_{,z} e^{4 \Psi}  +\frac{1}{r}(r^2e^{-2 \Psi} -{\cal W}^2 e^{2
\Psi}) A_{t,r} A_{t,z} \right .   \nonumber  \\
&& \left . \ - \frac{1}{r}A_{\varphi,r} A_{\varphi,z} e^{2 \Psi} 
+ \frac{1}{r}{\cal W}(A_{\varphi,r} A_{t,z}+ A_{\varphi,z}
A_{t,r})e^{2 \Psi}  \right ]. 
\label{eq:liga3}\end{eqnarray}
Finally, using the Einstein-Maxwell equation (\ref{eq:lamz}) follows
immediately that $F(v_+,v_-)$ vanishes and therefore the electrogeodesic
velocities satisfy the  constraint (\ref{eq:liga}) and so, if the
electrogeodesic equation admits solutions corresponding to circular orbits, we
have a well defined CRM. 


Another  possibility is to take the two
counterrotating fluids not circulating along  electrogeodesics but with equal
and opposite tangential  velocities,
\begin{equation}
v_{\pm} = \pm v = \pm \sqrt{p_\varphi / \epsilon}.
\end{equation}
This choice, that imply the existence of additional  interactions between  the
two streams (e.g. collisions), leads to a complete  determination of the
velocity vectors.  However, this can be made only when $0 \leq |p_\varphi/
\epsilon| \leq 1$.  In the general case, the two counterrotating streams
circulate with different velocities and we can write (\ref{eq:liga}) as 
\begin{equation}
 v_+ v_- = - \frac {p_\varphi}{\epsilon}.
\end{equation}
However, this relation does not determine completely the tangential velocities,
and therefore the CRM is undetermined.

In summary, the counterrotating tangential velocities can be explicitly
determined only if we assume some additional relationship between them, like
the equal and  opposite condition or the electro-geodesic condition. Now, can
happen that the obtained  solutions do not satisfy any of these two conditions.
That is, the counterrotating velocities are, in general, not completely
determined by the constraint (\ref{eq:liga}). Thus, the CRM is in general
undetermined since the counterrotaing energy densities and pressures can not be
explicitly written without a knowledge of the counterrotating tangential
velocities.


\section{Disks from a Kerr-Newman solution }

As an example of the above presented formalism, we consider the thin disk
models obtained by means of the ``displace, cut and reflect'' method applied to
the well known Kerr-Newman solution, which can be written as 
\begin{subequations}
\begin{eqnarray}
\Psi &=& \frac 12 \ln \left [ \frac {a^2x^2+b^2y^2 -c^2} {(ax+c^2)^2+b^2y^2}
\right ],  \\
\Lambda & = & \frac 12 \ln \left [ \frac {a^2x^2+b^2y^2  -c^2}{a^2(x^2-y^2)}
\right ],  \\
{\cal W }  &=&  \frac {c^2kb(1-y^2)(2ax +1 +   c^2) } {a(a^2x^2+b^2y^2
-c^2)},              \\
A_t  &=&  \frac {c \sqrt {2 (c^2-1)}(ax+c^2)}{(ax +c^2)^2+b^2y^2}, \\   
A_\varphi  & = & - k \frac b a (1-y^2) A_t, 
\end{eqnarray}\label{eq:metkn}\end{subequations}
where $a^2+b^2 = c^2 \geq 1$,  with
\begin{equation}
a=\frac{k}{m(1-qq*)}, \quad b=\frac{L}{m(1-qq*)}, \quad c=
\frac{1}{\sqrt{1-qq*}}, \quad k=\sqrt{m^2-L^2-e^2}, \quad |q|= \frac e m,
\end{equation}
where $m$, $L$ and $e$ are the mass, angular momentum and electric charge
parameters of the Kerr-Newman black hole,  respectively. Note that $c$ is the
parameter   that controls the electromagnetic field.  The prolate  spheroidal
coordinates, $x$ and $y$, are related with  the Weyl coordinates by 
\begin{equation}
r^2   = k^2 (x^2-1)(1-y^2),  \quad  \quad z + z_0 = kxy,  \label{eq:coorp}
\end{equation}
where  $1 \leq x \leq \infty$,  $0  \leq y \leq 1 $, and $k$ is an arbitrary
constant. Note that we have displaced the origin of the $z$ axis in $z_0$.  
This solution can be generated, in these coordinates,  using  the well-known 
complex potential formalism proposed by Ernst \cite{E2}  from the  Kerr vacuum
solution \cite{KSHM}.  When $c=1$ this solution reduces to the Kerr vacuum
solution.

Let be $\tilde D = k^2 D$, $\tilde T = k T$, and  $\tilde {\mbox{\sl 
j}}_t=k\mbox{\sl  j}_t$,  therefore
\begin{subequations} \begin{eqnarray}
\tilde T &=& \frac { 4c^2 a \bar y \{ 2 \bar x(1-\bar y^2) (\bar x^2+2a \bar
x+c^2) -(\bar x^2-\bar y^2)[a(\bar x ^2+1)+\bar x(1+c^2)] \} }{( \bar x^2-\bar
y^2)^{3/2}  [(a \bar x+c^2)^2 + b^2 \bar y^2]^{3/2}} ,  \\
&&    \nonumber   \\
\tilde D &=& \frac {16 c^4 a^2 \bar y^2 \{ [a(\bar x^2 +1)+ \bar x(1+c^2)]^2 - b^2 \tilde r^2 \}} {(\bar x^2 - \bar
y^2)[(a\bar x+c^2)^2 + b^2 \bar y^2]^3}, \\
&&    \nonumber   \\
\tilde {\mbox{\sl  j}}_t  &=&   \frac { 2 c \sqrt{2(c^2-1)} a\bar y \{ -b^2
\bar y^2(3a \bar x^2+2\bar x c^2-a) + (a \bar x+c^2)(a^2 \bar x^3+ac^2  \bar
x^2} {(\bar x^2- \bar y^2)^{1/2}[(a\bar
x+c^2)^2+b^2 \bar y^2]^{5/2}}  \nonumber  \\
&& -a^2 \bar x+2b^2 \bar x-ac^2) \},      \\
\mbox{\sl  j}_\varphi & = &  -\frac {2 c \sqrt{2(c^2-1)} b \bar y (1-\bar y^2)
\{-ab^2 \bar y^2(\bar x^2-1)+ (a \bar x+c^2)(3a^2 \bar x^3 } {( \bar x^2- \bar y^2)^{1/2}[(a \bar
x+c^2)^2+b^2 \bar y^2] ^{5/2}}  \nonumber  \\
&&  +5ac^2  \bar x^2-a^2 \bar x+2b^2 \bar x+2c^4 \bar x-ac^2) \}.
\end{eqnarray} \end{subequations}
In the above expressions $\bar{x}$ and  $\bar{y}$ are given by
\begin{subequations}\begin{eqnarray}
2\bar{x}  & = \sqrt {\tilde{r}^2 + (\alpha + 1)^2} + \sqrt {\tilde{r}^2 +
(\alpha - 1)^2}, \label{eq:xbar} \\
2\bar{y}  & = \sqrt {\tilde{r}^2 + (\alpha + 1)^2} - \sqrt  {\tilde{r}^2 +
(\alpha - 1)^2},\label{eq:ybar}
\end{eqnarray}\end{subequations}
where $\tilde{r}=r/k$ and $\alpha = z_0/k$, with  $\alpha >1$.

Now, in order to analyze the behavior of $D$, is enough to consider the
expression
\begin{equation}
 \tilde D_0 =  [a(\bar x^2 +1)+ \bar x(1+c^2)]^2 - b^2 \tilde r^2,  
\end{equation}
that can be written as
\begin{equation} \begin{array}{rcl} 
\tilde D_0 &=& a(1+c^2)R_+ [\alpha(\alpha-1)+ 2 + \tilde r ] + a(1+c^2)R_-
[\alpha(\alpha+1) + 2 + \tilde r ]   \\  
&&    \\
&& + \frac 12 R_+ R_- [(c^2+1)^2 + a^2( \tilde r^2 +  \alpha ^2+3)]   + \frac
12 [(c^2+1)^2 + 5a^2] \\ 
&&                       \\
&&+\frac 12 \tilde r^2 [c^4+1+a^2 (\tilde r^2+2\alpha^2+6)] +\frac 12
\alpha^2[(c^2+1)^2 + a^2 (\alpha^2 + 2 )] , \label{eq:d0}
\end{array}\end{equation}
where $R_\pm = \sqrt {\tilde{r}^2 + (\alpha \pm 1)^2}$. Since $\alpha(\alpha \mp 1)+ 2 >0$ for any $\alpha$, from (\ref{eq:d0}) follows that $D$  always is  a positive quantity for
Kerr-Newman fields and therefore the eigenvalues of the energy-momentum tensor
are always real quantities. So we conclude that these disks can be interpreted,
for all the values of parameters, as a matter distribution with currents and
purely azimuthal pressure and without heat flow.

We can see also that for the vacuum case, when $c = 1$, $D$  is everywhere
positive.  These disks, obtained from the Kerr  vacum solution, were previously
considered by Gonz\'alez and  Letelier in the reference \cite{GL2}. In this
previous work, due to a mistake in the computation of the expressions for
$\tilde T$ and $\tilde D$, was concluded that the  energy-momentum tensor could
present complex eigenvalues for some values of the parameters. As we can see
from the expressions presented here, this is not correct and, in all the cases,
we have a matter distributions with purely azimuthal pressure and without heat
flow for all the values of parameters.

In order to study the behavior of the main physical  quantities associated with
the disks, we  perform a graphical analysis of them for disks  with  $\alpha =
2$, $b=0.2$ and $c=1.0$, $1.5$,  $2.0$, $2.5$, and $3.5$, as functions of
$\tilde r$.   For these values of the parameters we find that $\Psi_{,z}$ is a
positive quantity in  agreement with the strong energy condition. Therefore,
$\epsilon=-\lambda_- $ and $p_\varphi = \lambda_+ $. However, one also finds
values of the parameter for which  $\Psi_{,z}$ takes negative values.
Furthermore, we take $\nu = -1$ in order to $V^a$ be a future-oriented timelike
vector.

\begin{figure*}
$$
\begin{array}{cc}
\tilde \epsilon & \tilde p_\varphi  \\   
\epsfig{width=3in,file=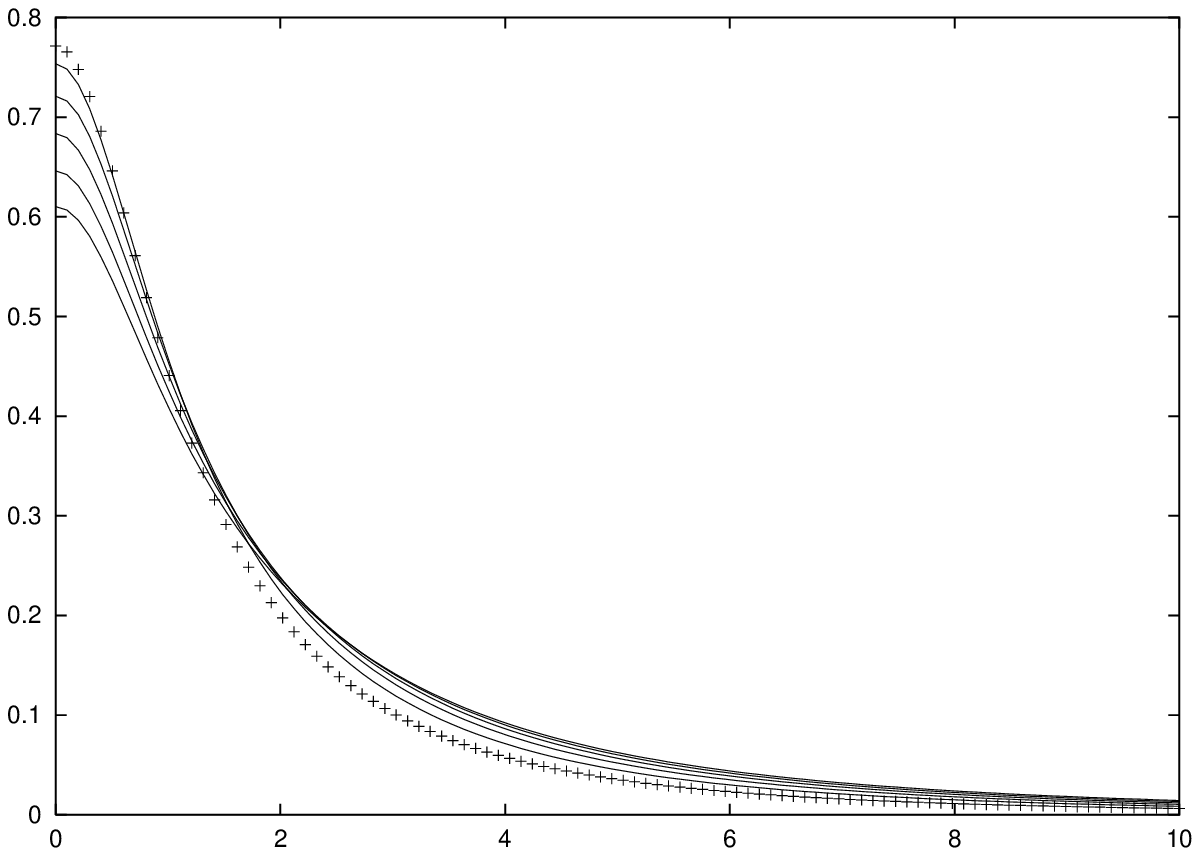} & \epsfig{width=3in,
file=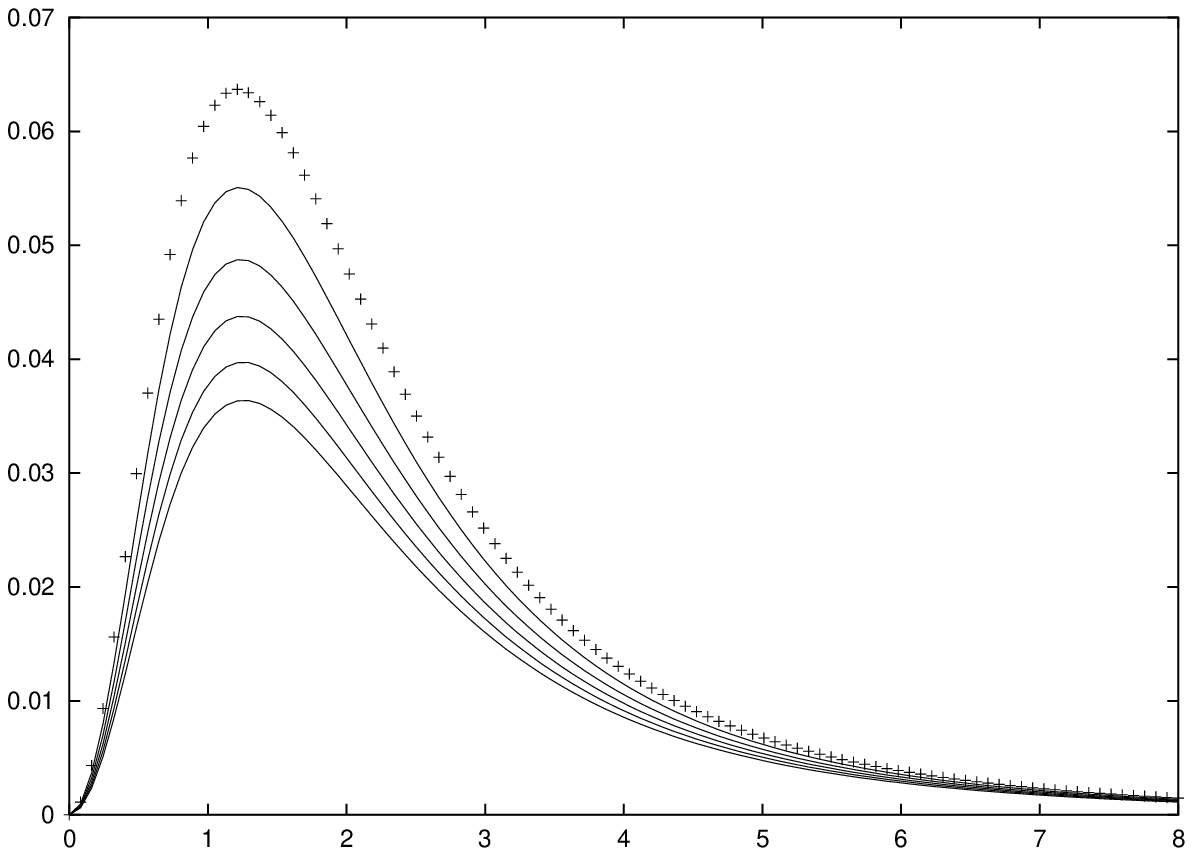} \\
\tilde r & \tilde r    \\
(a)     &   (b)
\end{array}
$$	
\caption{ $(a)$ The surface energy density $\tilde \epsilon$
and $(b)$ the azimuthal pressure $\tilde p_\varphi$ for Kerr-Newman disks with  $\alpha = 2$,
$b=0.2$ and  $c=1.0$ (curves with crosses), $1.5$, $2.0$, $2.5$, $3.0$, and
$3.5$ (bottom curves), as  functions of $\tilde r$.} \label{fig:enprkn}
\end{figure*}

In Fig. \ref{fig:enprkn} we  show the surface energy density  $\tilde
\epsilon$  and the  azimuthal pressure $\tilde p_\varphi$. We see  that the
energy density presents a maximum at $\tilde r = 0$ and then decreases rapidly
with $\tilde r$, being always a positive quantity in agreement with the weak
energy condition. We also see that  the presence of electromagnetic  field
decreases the  energy density  at the central region of the disk and  later
increases it.  We  can observe that the pressure increases rapidly as one moves
away from the disk center, reaches a maximum and  later  decreases rapidly. We
also observe  that the electromagnetic field decreases the pressure everywhere
on the disk.

\begin{figure*}
$$
\begin{array}{cc}
\tilde { \mbox{\sl j} }_t    & - \mbox{\sl j}_\varphi   \\
\epsfig{width=3in,file=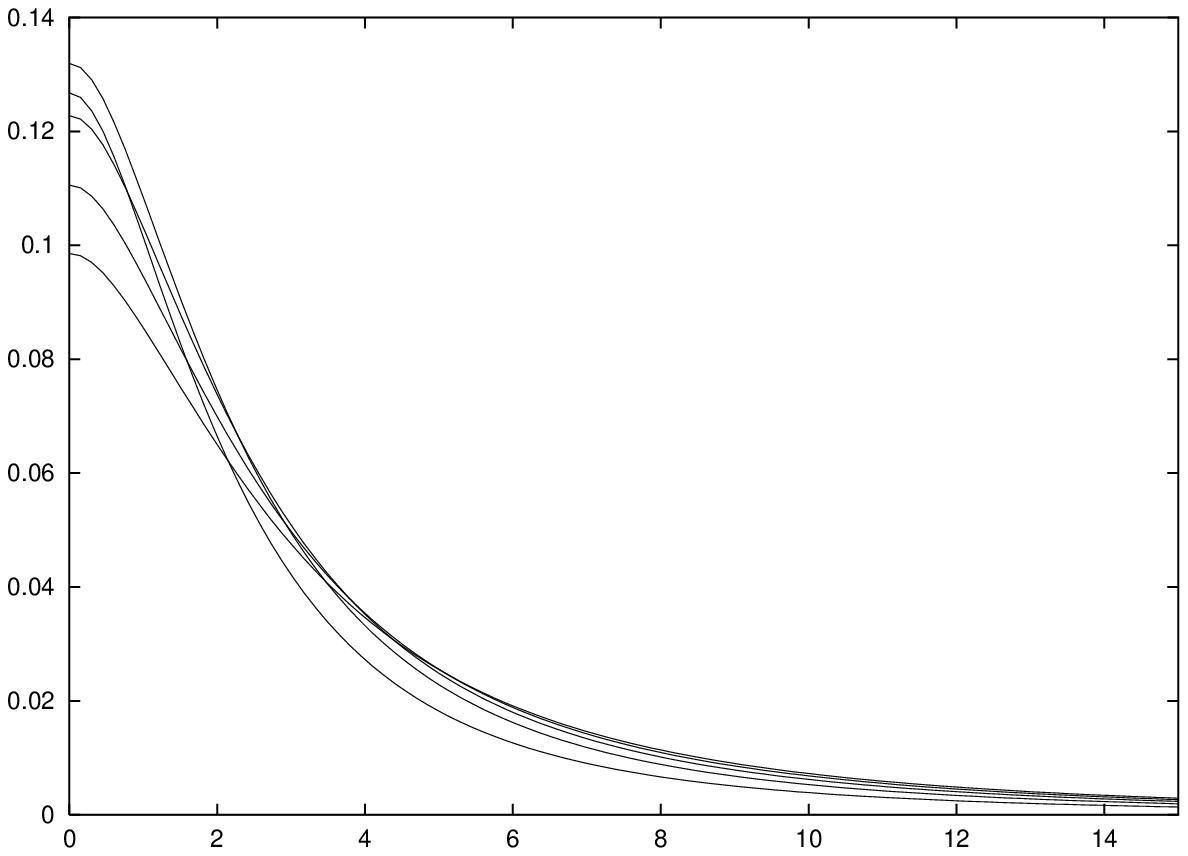} & \epsfig{width=3in,
file=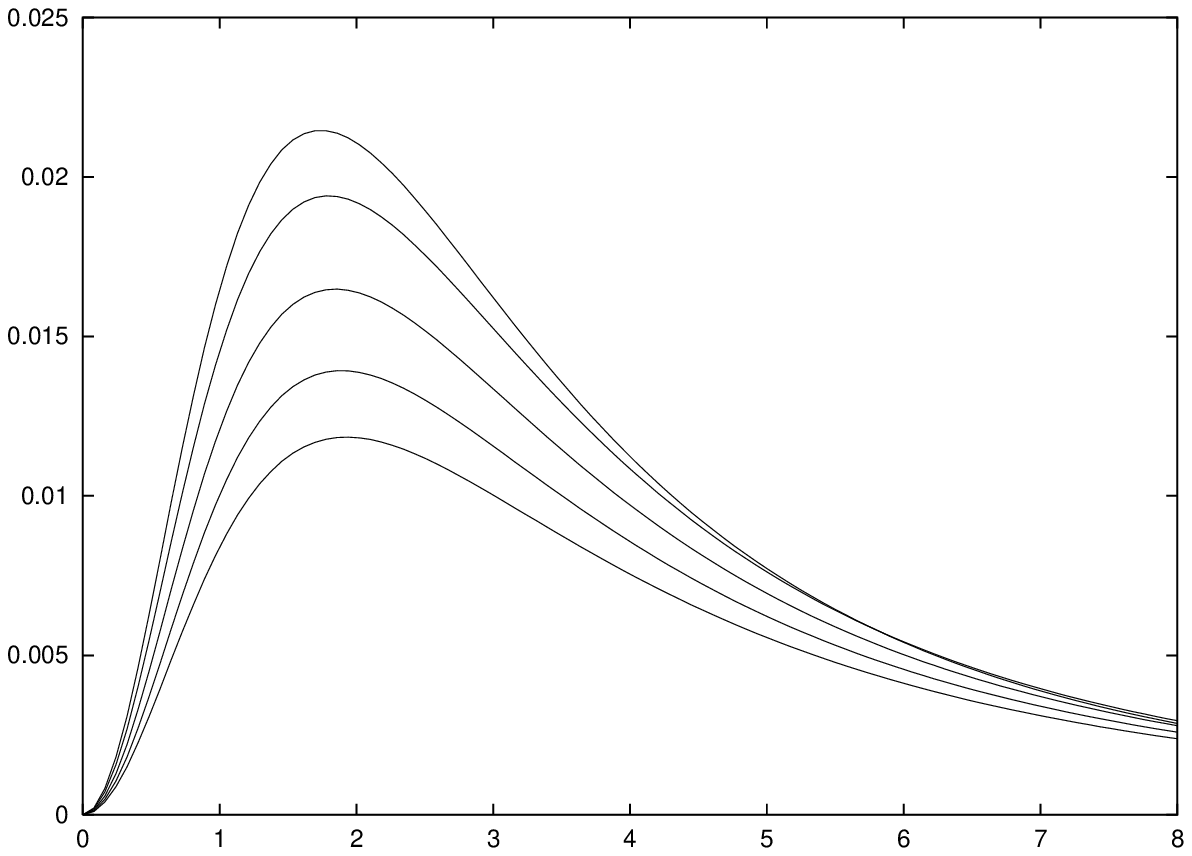} \\
\tilde r & \tilde r    \\
(a)     &   (b)
\end{array}
$$	
\caption{$(a)$ The surface electric charge density  
$\tilde {\mbox{\sl j} }_t$ and $(b)$ the azimuthal current
density  $\mbox{\sl j}_\varphi$ for Kerr-Newman disks 
with $\alpha = 2$, $b=0.2$ and
$c=1.0$ (axis $\tilde r$), $1.5$ (top
curves), $2.0$, $2.5$, $3.0$, and $3.5$ (bottom curves), as functions of $\tilde
r$.} \label{fig:j0j1kn}
\end{figure*}

\begin{figure*}
$$
\begin{array}{cc}
-\tilde { \mbox{\sl j} }^{\hat 0}    & 
-\tilde { \mbox{\sl j} }^{\hat 1}   \\
\epsfig{width=3in,file=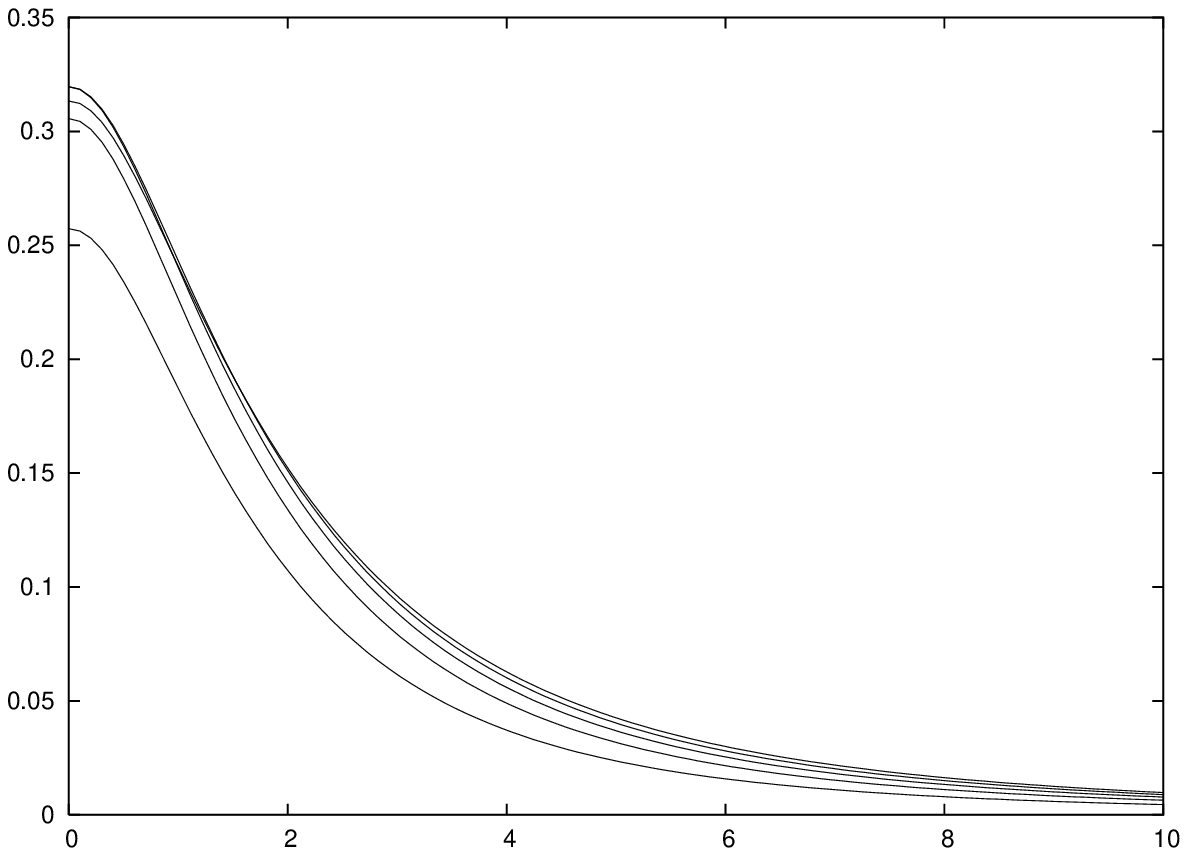} & 
\epsfig{width=3in,file=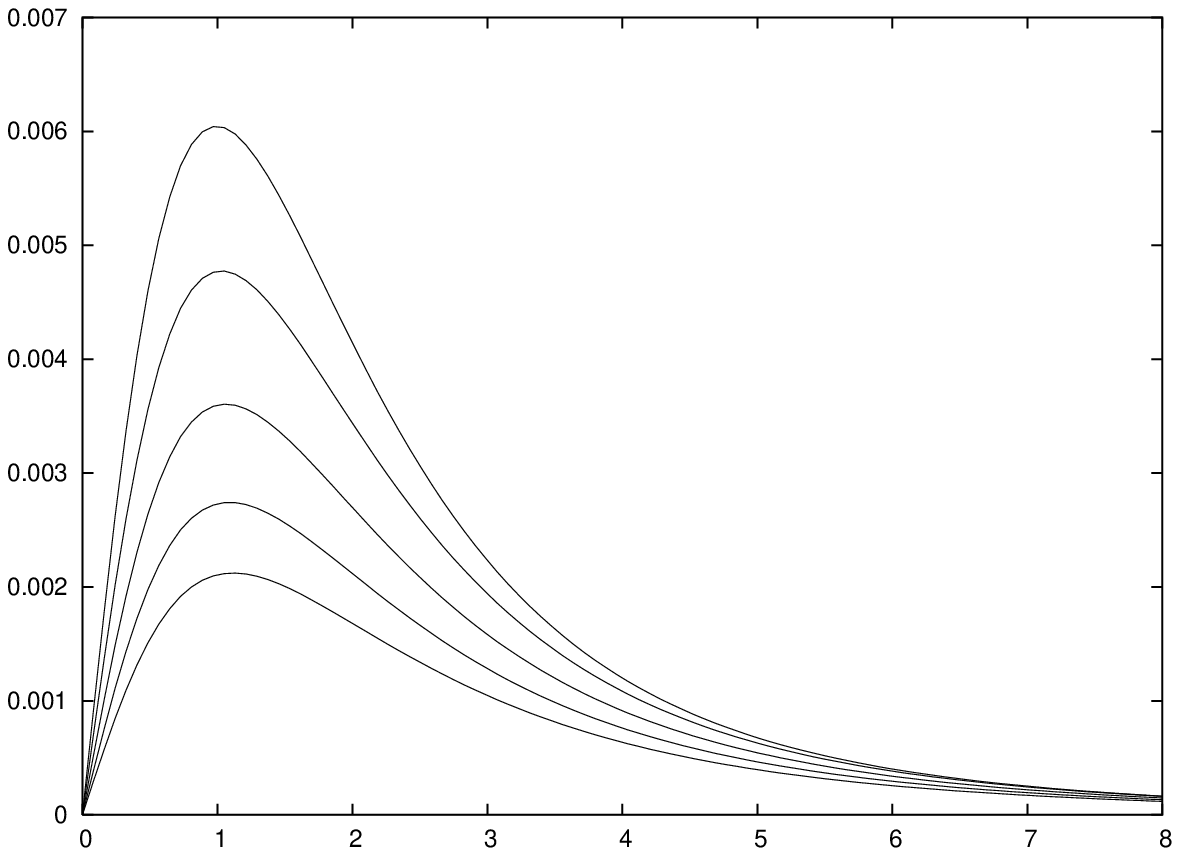} \\
\tilde r & \tilde r    \\
(a)     &   (b)
\end{array}
$$	
\caption{For Kerr-Newman disks we plot, as function 
of $\tilde r$, $(a)$  $ \tilde { \mbox{\sl j} }^{\hat 0}$
with  $\alpha = 2$,  $b=0.2$, and
$c=1.0$ (axis $\tilde r$), $1.5$  
(bottom curve, away from the center of disk),
$2.0$, $2.5$, $3.0$, and $3.5$ (top curve, away from  the center of disk), and 
$(b)$  $\mbox{\sl j}^{\hat 1}$ 
also with  $\alpha = 2$,  $b=0.2$,
and $c=1.0$  (axis $\tilde r$), $1.5$ (top curve),
$2.0$,  $2.5$, $3.0$, and $3.5$ (bottom curve).
}\label{fig:sijkn}
\end{figure*}

The electric charge density $ \tilde {\mbox{\sl j}}_t $ and  the  azimuthal
current density $\mbox{\sl j}_\varphi$, measured  in the coordinates frame, 
are  represented in Fig.  \ref{fig:j0j1kn}, whereas the electric charge density
$\tilde { \mbox{\sl j} }^{\hat 0}$ and  the azimuthal current density $\tilde {
\mbox{\sl j} }^{\hat 1}$,  measured by the comoving observer, are  represented
in Fig. \ref{fig:sijkn}. We observe that the electric charge density has a 
similar behavior to the energy and that the current density have a similar
behavior to the  pressure which is consistent with the fact that the mass  is
more concentrated in the disks center. We also computed  this functions  for
other values of the parameters and, in  all the cases, we found  the same
behavior. 

\begin{figure*}
$$
\begin{array}{cc}
v _+    & -v_ - \\   
\epsfig{width=3in,file=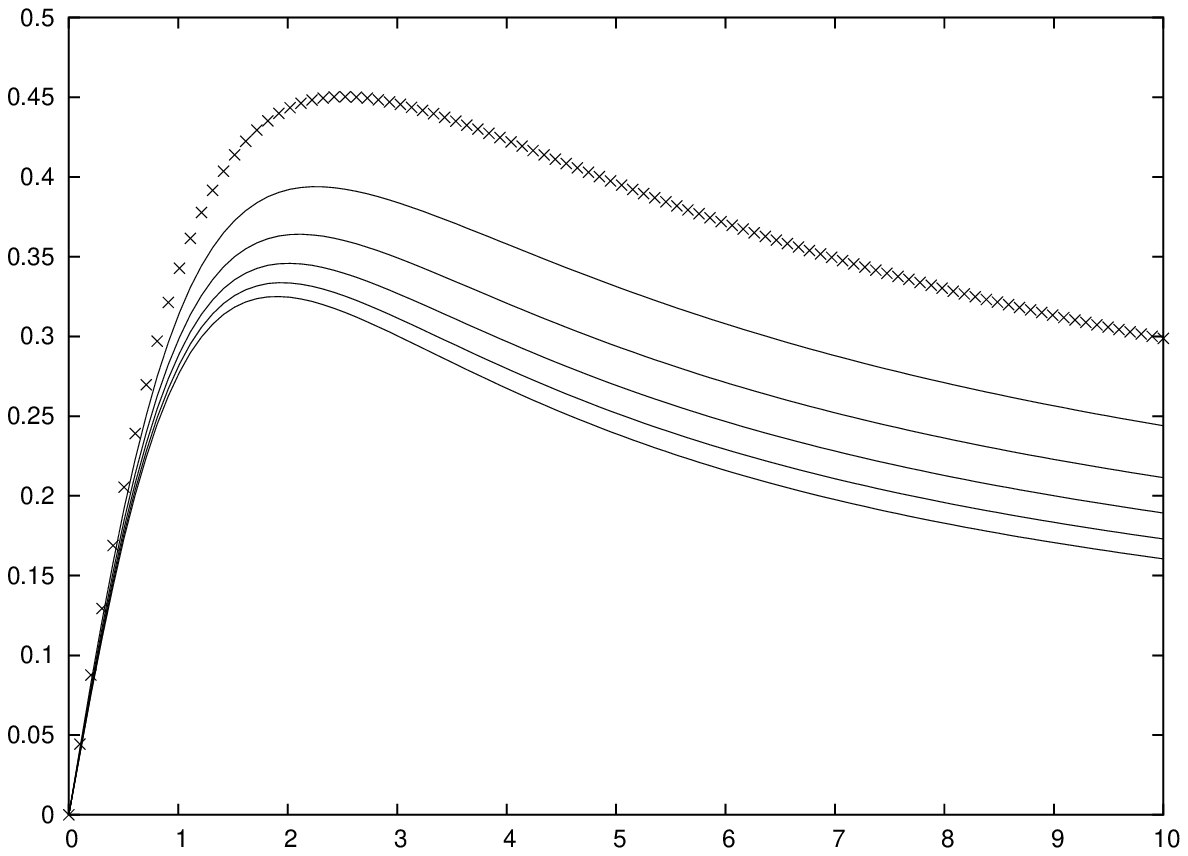} & \epsfig{width=3in,
file=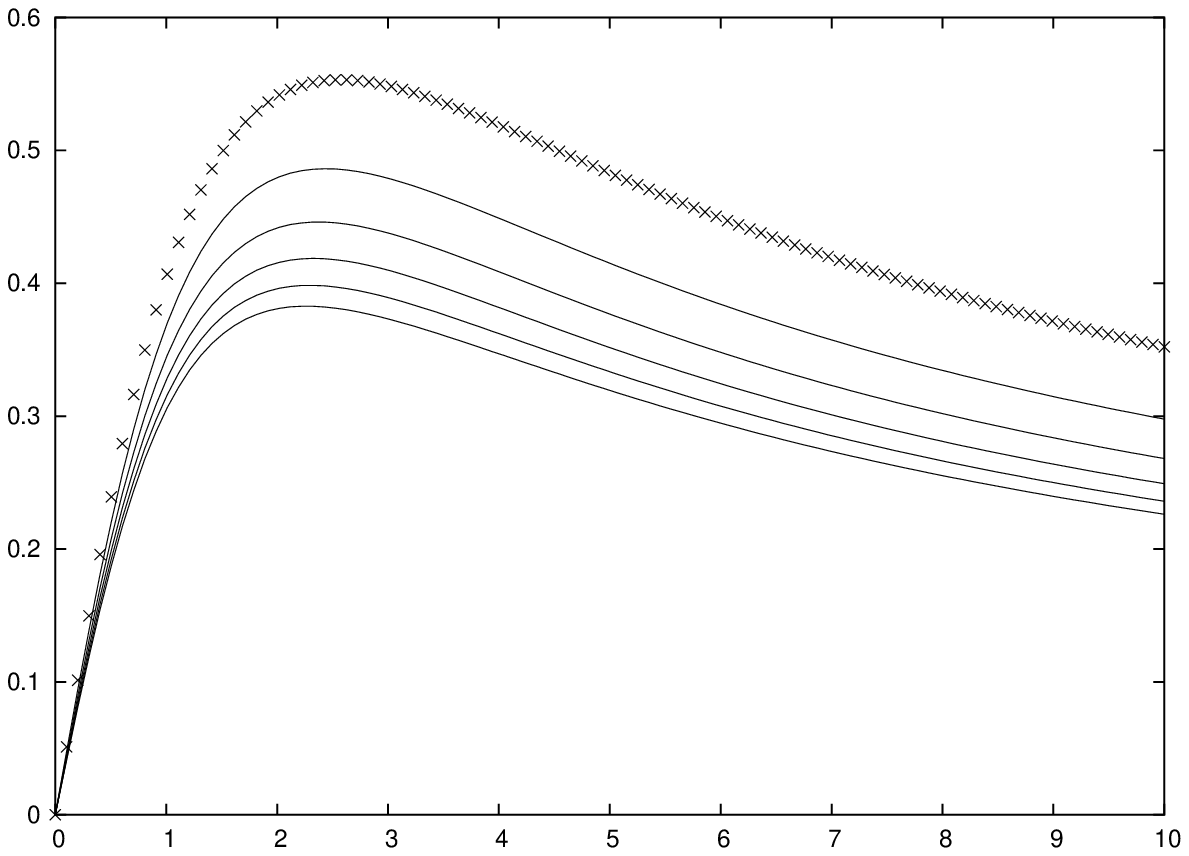}    \\
\tilde r & \tilde r    \\
(a)     &   (b)
\end{array}
$$	
\caption{The tangential velocities $(a)$ $v _+$ and $(b)$ $v _-$  for electrogeodesic Kerr-Newman disks
with  $\alpha = 2$, $b=0.2$ and  $c=1.0$ (curves with crosses), $1.5$, $2.0$,
$2.5$, $3.0$, and $3.5$ (bottom curves), as  functions of $\tilde r$.}
\label{fig:velkn}
\end{figure*}

\begin{figure*}
$$
\begin{array}{cc}
h _+^2    & h_-^2  \\   
\epsfig{width=3in,file=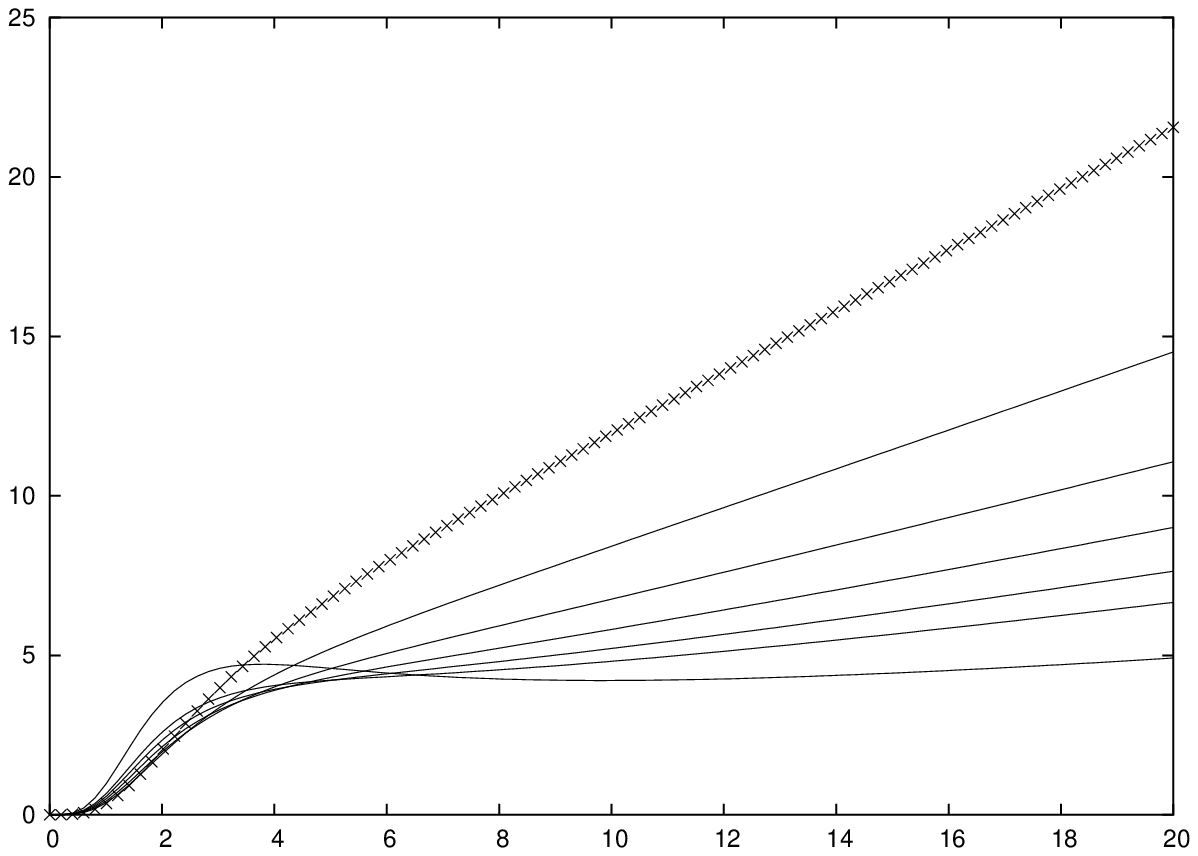} & \epsfig{width=3in,
file=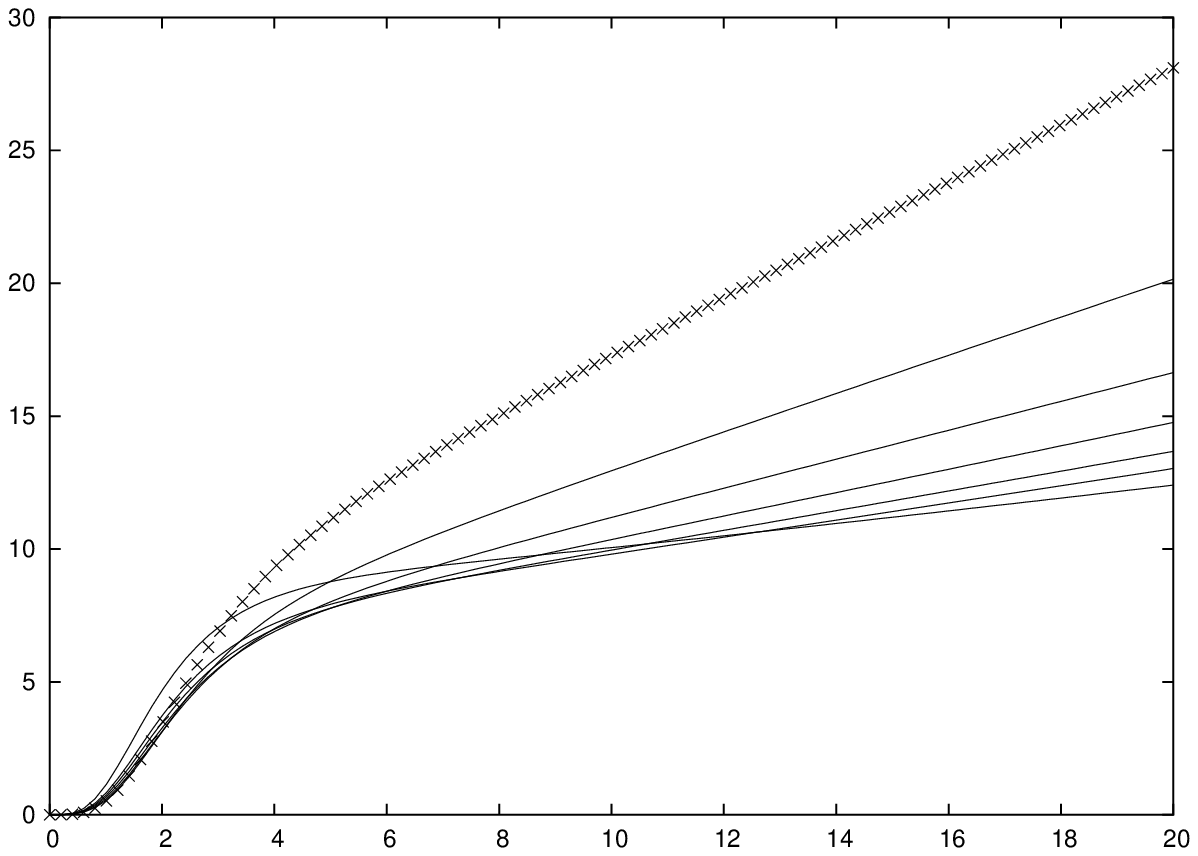}    \\
\tilde r & \tilde r    \\
(a)     &   (b)
\end{array}
$$	
\caption{The specific angular momenta $(a)$ $h _+^2$ and (b) $h_-^2$  for electrogeodesic Kerr-Newman disks
with  $\alpha = 2$, $b=0.2$ and  $c=1.0$ (curves with crosses), $1.5$, $2.0$,
$2.5$, $3.0$, $3.5$, and $6.0$ (bottom curves), as functions of $\tilde r$.}
\label{fig:h2elkn}
\end{figure*}

\begin{figure*}
$$
\begin{array}{cc}
\tilde \epsilon _+    & \tilde \epsilon_ - \\   
\epsfig{width=3in,file=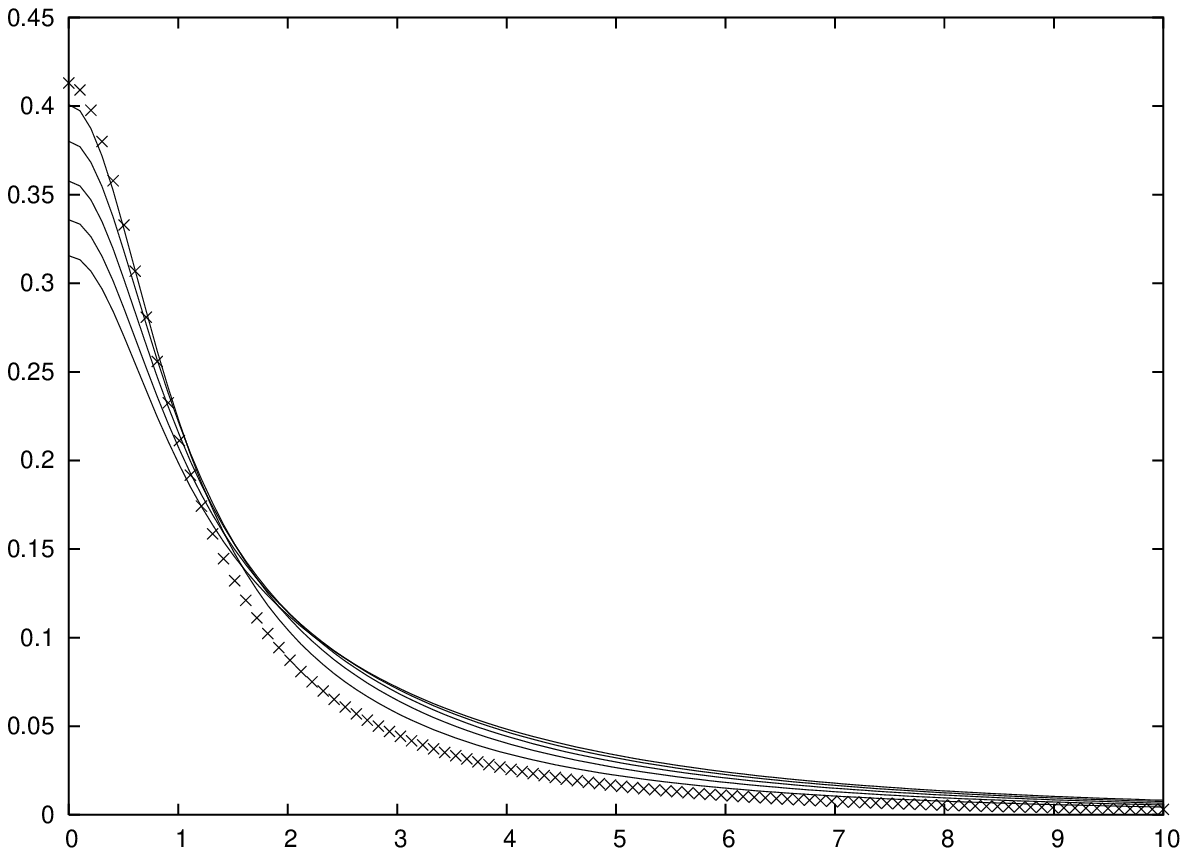} & \epsfig{width=3in,
file=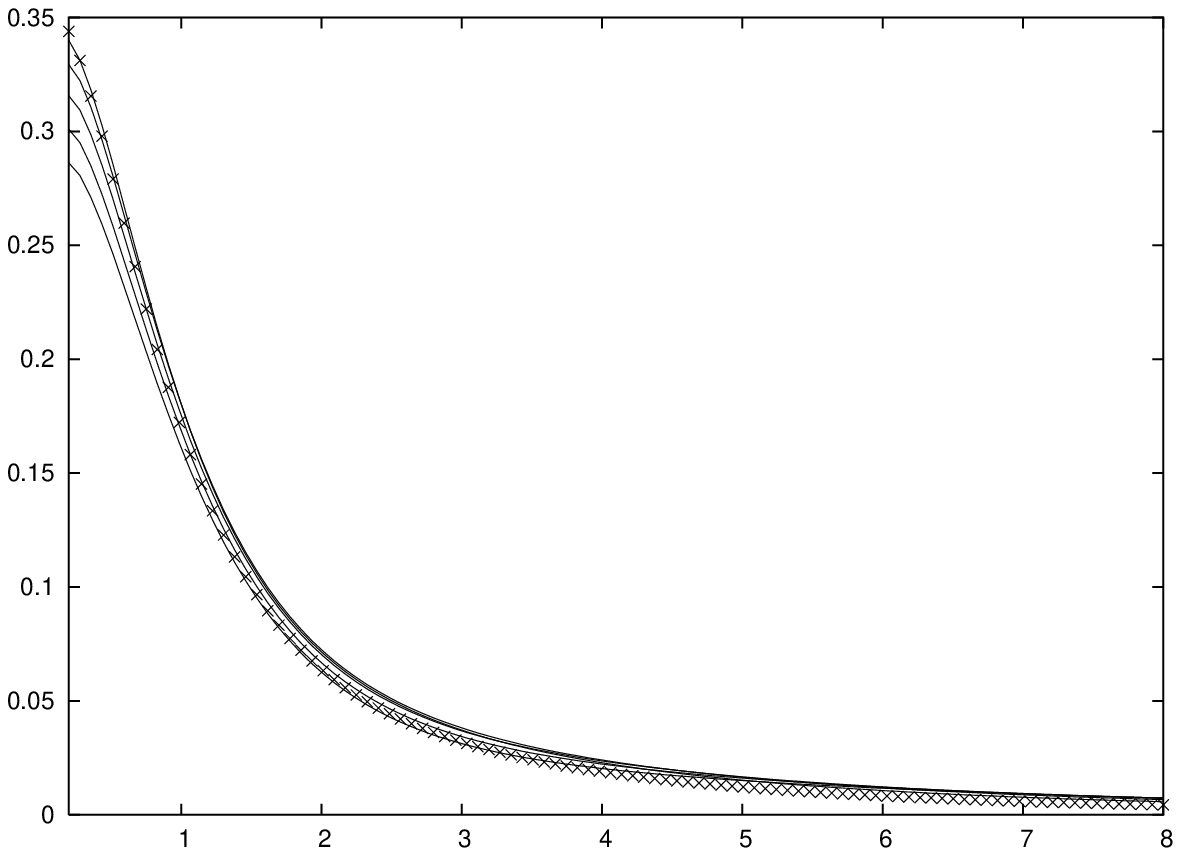}    \\
\tilde r & \tilde r    \\
(a)     &   (b)
\end{array}
$$	
\caption{The surface energy densities $(a)$ $\tilde \epsilon _+$ and $(b)$ $\tilde \epsilon _-$  for
electrogeodesic Kerr-Newman disks with  $\alpha = 2$, $b=0.2$ and  $c=1.0$
(curves with crosses), $1.5$, $2.0$, $2.5$, $3.0$, and $3.5$ (bottom curves), as  functions of $\tilde r$.} \label{fig:enelkn}
\end{figure*}

\begin{figure*}
$$
\begin{array}{cc}
-\tilde \sigma _+    & -\tilde \sigma _ - \\   
\epsfig{width=3in,file=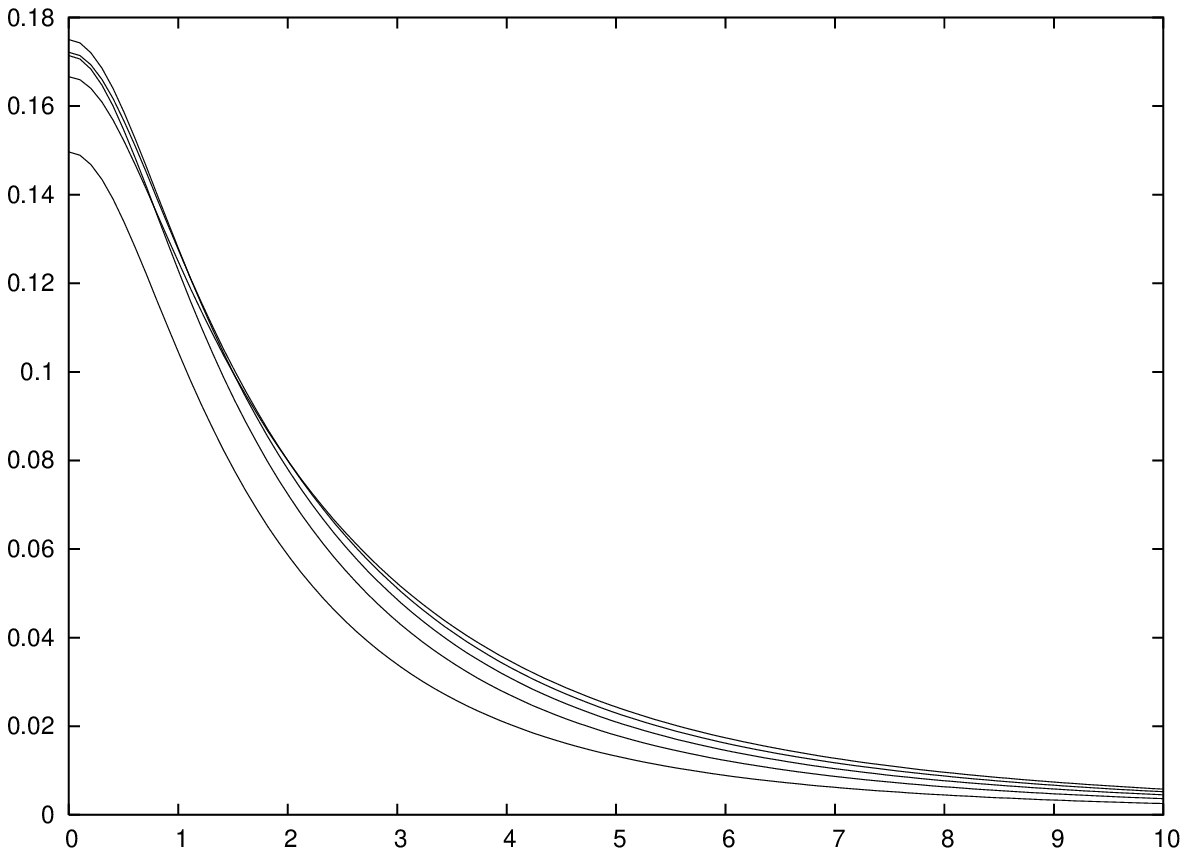} & \epsfig{width=3in,
file=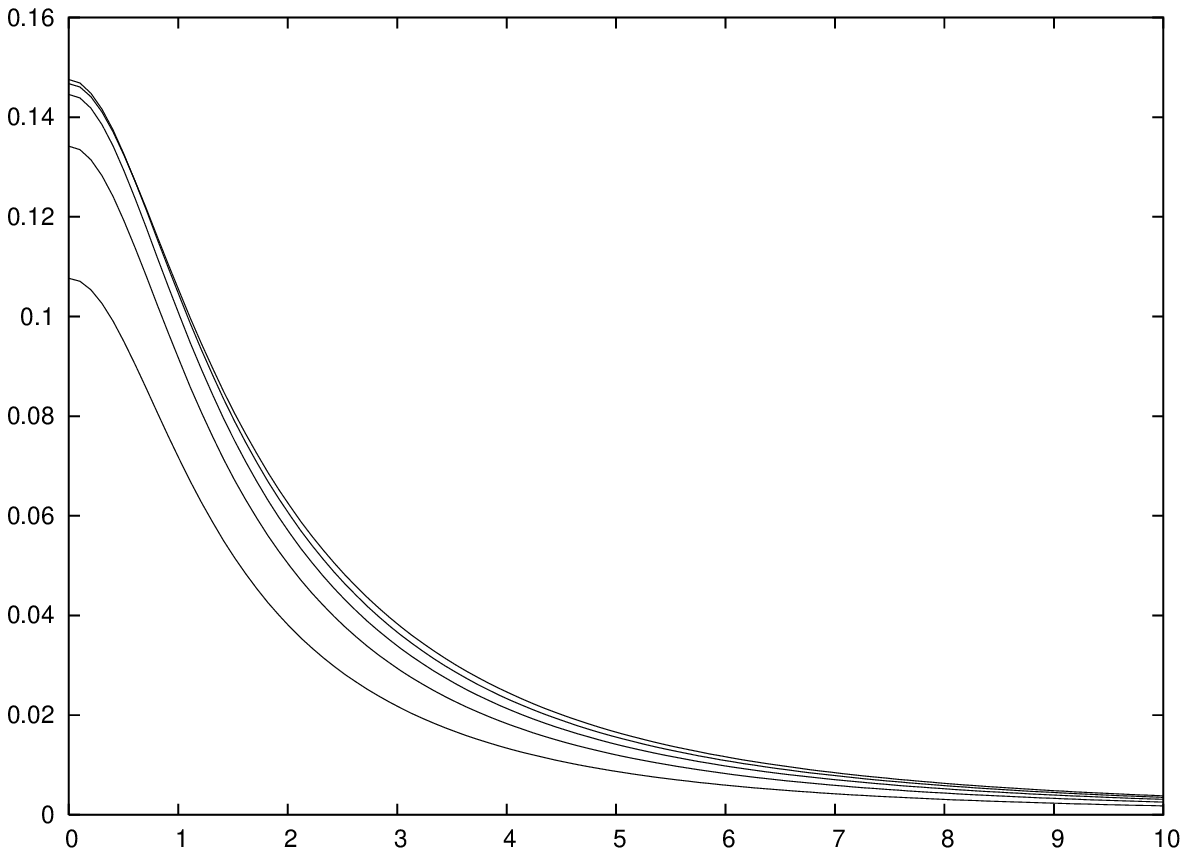}    \\
\tilde r & \tilde r    \\
(a)     &   (b)
\end{array}
$$	
\caption{The surface electric charge densities $(a)$ $\tilde \sigma _+$ and $(b)$ $\tilde \sigma _-$  for
electrogeodesic Kerr-Newman disks with  $\alpha = 2$, $b=0.2$ and  $c=1.0$
(axis $\tilde r$), $1.5$, $2.0$, $2.5$, $3.0$, and $3.5$ (top curves away from
the center of the disk), as functions of $\tilde r$.} \label{fig:sielkn}
\end{figure*}

We now consider the CRM for the same values of the parameters. We first
consider the two counterrotating streams circulating along electrogeodesics. In
Fig. \ref{fig:velkn} we plot the tangential velocity curves, $v_+$ and $v_-$.
We see that these  velocities  are always less than the light velocity. We also
see that the inclusion of the electromagnetic field make less relativistic
these disks.  In Fig.  \ref{fig:h2elkn} we have  drawn the  specific angular
momenta  $h_+^2$ and $h_-^2$  for the same values  of the parameters. We see 
that the presence of  electromagnetic field can make unstable these orbits 
against radial  perturbations. Thus the CRM cannot apply for $c=6$ (bottom
curve).  In Fig. \ref{fig:enelkn}  we have  plotted the surface energy
densities $\tilde \epsilon_+$ and  $\tilde \epsilon_-$. We see that these
quantities have a similar behavior to the energy density $\tilde \epsilon$.  
In Fig. \ref{fig:sielkn}, we plotted the surface electric charge densities
$\tilde \sigma _+$ and  $\tilde \sigma _-$.  We find that these quantities have
also a similar behavior to  $\tilde \epsilon_\pm $. 

The Figs. \ref{fig:velkn} - \ref{fig:sielkn} show that the two fluids
are continuous in $r$ which implies to have  two particles in counterrotating
movement in the same point in spacetime. So this model could be possible when
the distance between streams (or between the counterrotating particles) were
very small in comparing with the length  $r$  so that we can consider, in
principle, the fluids continuous  like is the case of  counterrotating gas
disks present  in disk galaxies.

Finally, in the case when  the two
fluids move with equal and opposite tangential velocities (non-electrogeodesic
motion) we find that the physical quantities have a similar  behavior to the
previous one. 

\section{Discussion}

We presented a detailed analysis of the energy-momentum tensor and the surface
current density for electrovacuum stationary axially symmetric relativistic
thin disks of infinite extension  without radial stress, in the case when the
energy-momentum tensor of the disks can be diagonalized, so that there is not
heat flow. The surface energy-momentum tensor and the surface current density
were expressed in terms of the comoving tetrad and explicit expressions were
obtained for the kinematical and dynamical variables that characterize the
disks. That is, we obtained  expressions for the velocity vector of the disks,
as well for the energy density, azimuthal pressure, electric charge density and
azimuthal current density.

We also presented in this paper the stationary generalization of the 
Counterrotating Model (CRM) for electrovacuum thin disks  previously  analyzed
for the  static case in \cite{GG1,GG2}. Thus then, we were able to obtain
explicit expressions for all the quantities involved in the CRM that are
fulfilled when do not exists heat flow and when we do not have radial pressure.
We considered both counter rotation with equal and opposite velocities and
counter rotation along electrogeodesics and, in both of the cases, we found the
necessary conditions for the existence of a well defined CRM.

A general constraint over the counterrotating tangential velocities was
obtained, needed to cast the surface energy-momentum tensor of the disk in such
a way that can be interpreted as the superposition of two counterrotating dust
fluids. The constraint obtained is the generalization of the obtained for the
vacuum case in \cite{GL2}, for disks without radial pressure or heat flow,
where we only consider counterrotating fluids circulating along geodesics. We
also found that, in general, there is not possible to take the two
counterrotating tangential velocities as equal and opposite neither take the
two counterrotating fluids as circulating along geodesics.

A simple family of models of counterrotating charged disks based on the
Kerr-Newman solution were considered where we obtain some disks with a CRM well
behaved.  We also find that the disks constructed  from the Kerr-Newman
solution can  be interpreted,  for all the values of parameters, as a matter
distribution with currents and purely azimuthal pressure and without heat flow.
We obtain, for all the values of parameters, counterrotating Kerr-Newman disks
that are in  agreement with  all the energy conditions. Finally, the
generalization  of these models to the case of  electrovacuum stationary
axially symmetric solutions where the  energy-momentum tensor of the disk  can
to present complex eigenvalues for some values of the parameters, the stability of counterrotating fluids taking into account the collective behavior of the particles, and 
a thermodynamic analysis of the disks, will be considered in future works.    

\subsection*{Acknowledgments}

The authors want to thank the financial support  from   COLCIENCIAS, Colombia.

\newpage


\end{document}